\documentclass{article}
\usepackage{frascatiphys,here,graphicx,subfigure}
\input pubboard/babarsym

\begin{document}
\title{ 
Charmonium spectroscopy at \babar\
}
\author{
Arafat Gabareen Mokhtar        \\
{\em Dept. of Physics} \\ 
{\em Colorado State University}\\
{\em  Fort Collins, CO 80523}
{\em  USA }\\ \\
for the \babar\ Collaboration
}

\maketitle
\baselineskip=11.6pt
\begin{abstract}
 
The charmonium-like states, $Y(4260)$, $Y(4350)$, produced via initial
state radiation, as well as the $X(3872)$, and $Y(3940)$, produced in
$B$ meson decays from the \babar\ B-factory are reviewed. These mesons
do not seem consistent with conventional charmonium models, and
several alternate hypotheses have been proposed to explain these new
discoveries.
\end{abstract}
\baselineskip=14pt
\section{Introduction}
\begin{figure}[htbp]
    \begin{center}
      {\includegraphics[scale=0.25]{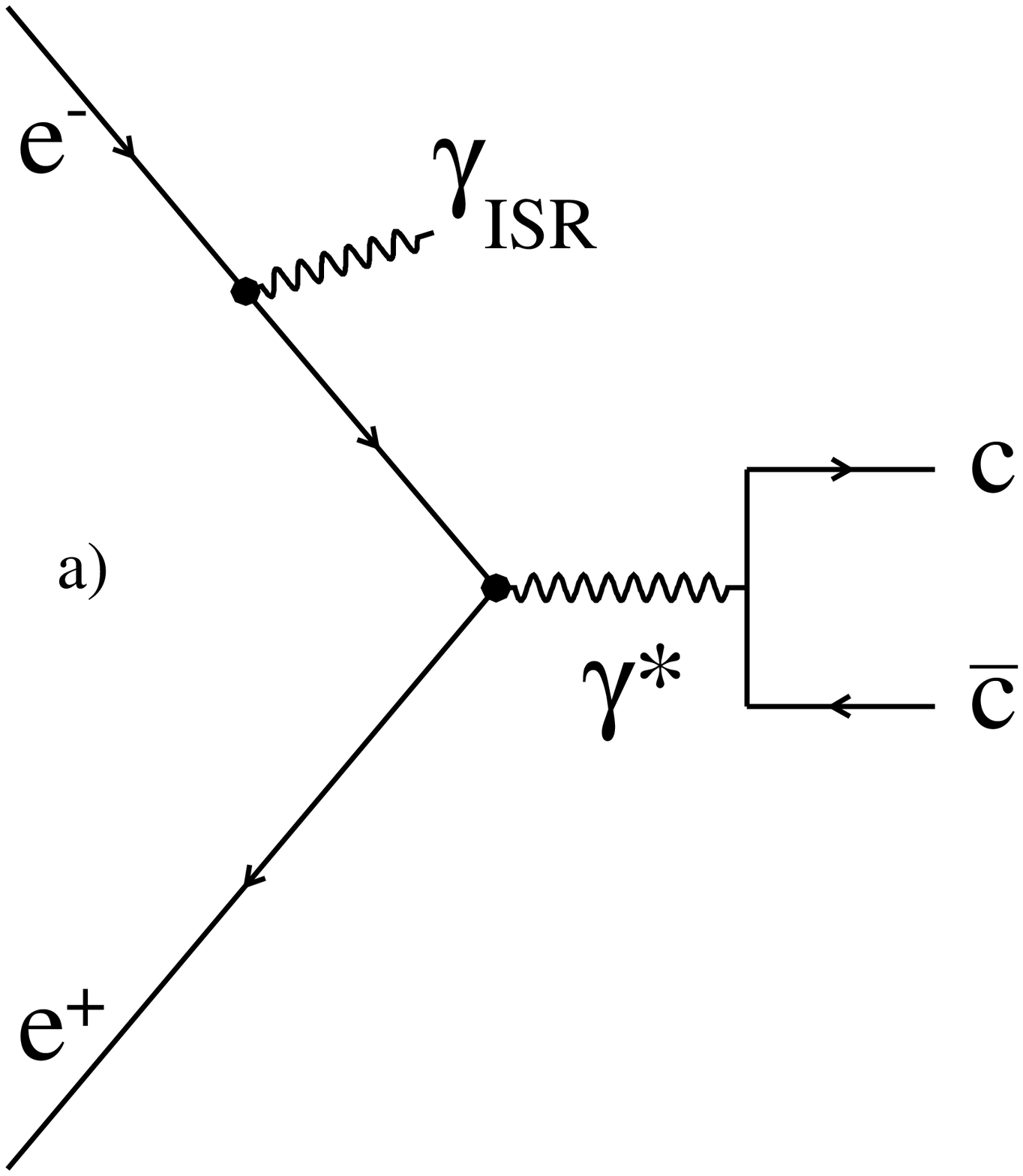}}
      {\includegraphics[scale=0.25]{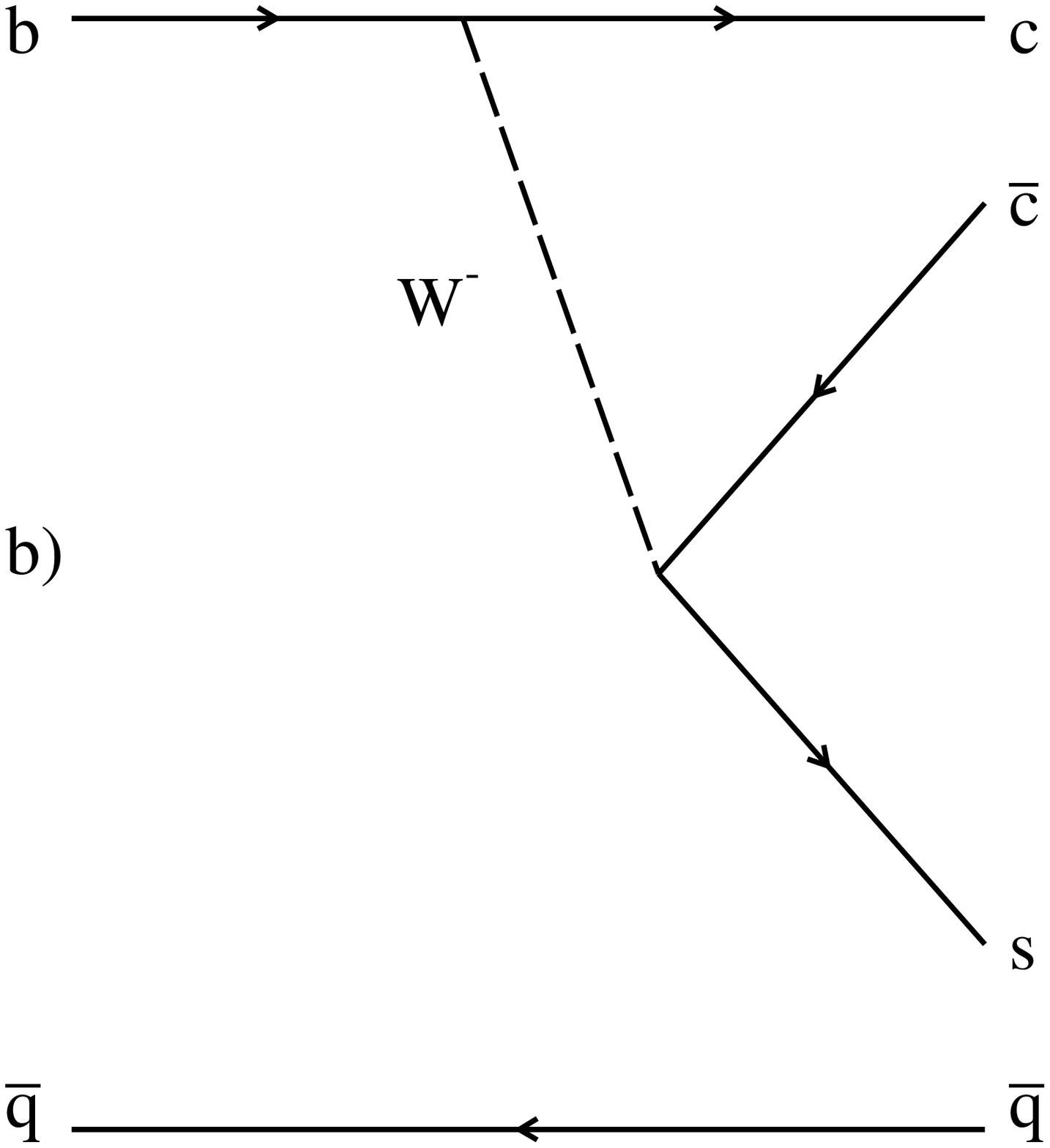}}
      \caption{\it Feynman diagrams representing charmonium production
      via a) ISR, b) $\bar{B}$ decay.}
      \label{diagrams}
    \end{center}
\end{figure}

Several charmonium-like states have been discovered recently at the
BELLE and \babar\ $B$-factories. These new states have been observed
in $e^+e^-$ initial state radiation (ISR) interactions or in
$B$-decays. The relevant Feynman diagrams representing ISR production
and $B$-decay are shown in Fig.~\ref{diagrams}. In ISR events, a real
photon is emitted from the incoming electron or positron and
subsequently the electron and positron annihilate to yield a virtual
photon ($\gamma^\star$) which couples to a $c\bar{c}$ system with
c.m. energy lower than the nominal value, and thus charmonia can be
produced. In $\bar{B}$-decay, a $W^-$ from the $b$-quark yields an
$s\bar{c}$ system; the $\bar{c}$ combines with the $c$ quark from $b$
decay to produce a charmonium state, while the $s$ quark and $\bar{q}$
spectator yield a strange meson.

The discovery of the charmonium-like state $X(3872)\rightarrow
J/\psi\pi^+\pi^-$ was reported by the BELLE
collaboration\cite{belle_x3872}, and confirmed by the
CDF\cite{cdf_x3872}, D0\cite{d0_x3872}, and \babar\/\cite{babar_x3872}
experiments. Since then several charmonium-like states have been
discovered at the $B$-factories, however they do not seem consistent
with conventional charmonium spectroscopy. Alternative explanations for
these states have been proposed, such as molecules, 4-quark states,
hybrids, etc.

In this report, we review briefly the latest results from the \babar\
experiment concerning four of these charmonium-like states.

\section{The $Y(4260)$}
\begin{figure}[htbp]
    \begin{center}
      {\includegraphics[scale=0.5]{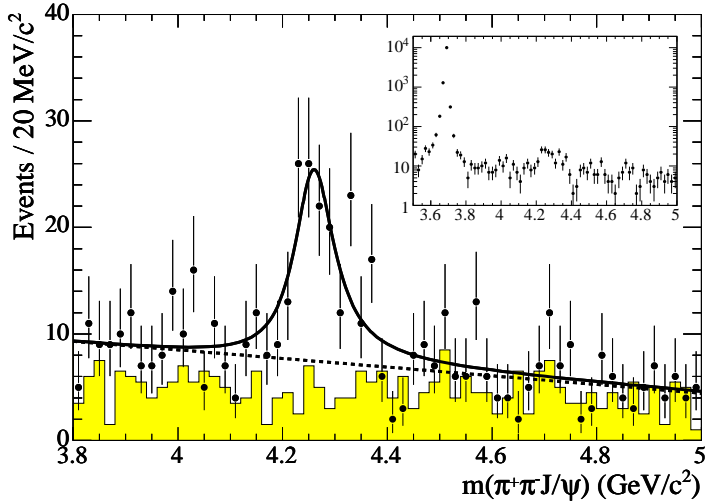}}
      \caption{\it The $J/\psi\pi^+\pi^-$ invariant mass distribution
      in the range $3.8-5.0$ GeV/c$^2$. The dots represent the data,
      the filled histogram shows the background from the $J/\psi$
      side-bands, the solid curve represents the fit result and the
      dashed line shows the background.}
      \label{y4260}
    \end{center}
\end{figure}

The $Y(4260)\rightarrow J/\psi\pi^+\pi^-$ was discovered by \babar\ in
the ISR reaction
$e^+e^-\rightarrow\gamma_{ISR}J/\psi\pi^+\pi^-$\cite{babar_y4260}
using 233 fb$^{-1}$ of data, where detection of the ISR photon was not
required. The $J/\psi\pi^+\pi^-$ mass distribution is shown in
Fig.~\ref{y4260}, where in the sub-figure a broader mass region shows
the peak due to $\psi(2S)\rightarrow J/\psi\pi^+\pi^-$; an enhancement
is observed at $\sim 4.26$ GeV/c$^2$. The mass region
$3.8<m_{J/\psi\pi^+\pi^-}<5$ GeV/c$^2$ is fitted with a Breit-Wigner
signal function and a second order polynomial background. The
background from $J/\psi$ side-band does not show any peaking
structure. The number of signal events extracted from the fit is
$125\pm 23$, the mass is $M_Y = 4259\pm8^{+2}_{-6}$ MeV/c$^2$, and the
width is $\Gamma_Y=88\pm23^{+6}_{-4}$ MeV. The branching
fraction obtained is $\Gamma_{Y,ee}*BF(Y(4260)\rightarrow
J/\psi\pi^+\pi^-)=5.5\pm 1.0^{+0.8}_{-0.7}$ eV. At \babar\/, no
evidence was found for the processes $Y(4260)\rightarrow
\phi\pi^+\pi^-$\cite{babar_y4260phipipi}, $Y(4260)\rightarrow
D\bar{D}$\cite{babar_y4260dd}, and $Y(4260)\rightarrow
p\bar{p}$\cite{babar_y4260pp}. A search for the $Y(4260)$ resonance
in $B$ decay was carried out, and a $3\sigma$ effect was
observed\cite{babar_y4260B}.

\section{The $Y(4350)$}
\begin{figure}[htbp]
    \begin{center}
      {\includegraphics[scale=0.9]{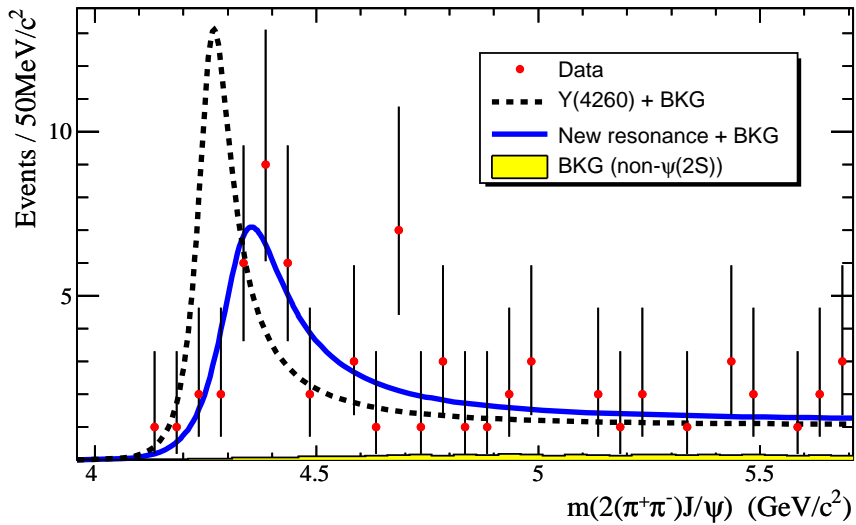}}
      \caption{\it The $2(\pi^+\pi^-)J/\psi$ invariant mass
      spectrum. The dots indicate the data and the shaded histogram
      represents the background. The solid curve shows the fit result
      with free mass and width parameters, while the dashed curve is
      obtained with the mass and width fixed to their
      $Y(4260)$ values\cite{babar_y4260}.}
      \label{y4350}
    \end{center}
\end{figure}

In \babar\/, a search for $Y(4260)\rightarrow \psi(2S)\pi^+\pi^-$
yielded instead evidence for a broad structure near $\sim 4.3$
GeV/c$^2$\cite{babar_y4350}. This enhancement is not consistent with
the $Y(4260)$ state. In Fig.~\ref{y4350}, the $2(\pi^+\pi^-)J/\psi$
invariant mass is shown for the data (dots) and for the background
(shaded histogram). The data points are fitted with a Breit-Wigner
signal function with fixed mass and width\cite{babar_y4260} (dashed
line), and again with mass and width as free parameters. The latter
fit yields mass $m=4324\pm 23$ MeV/c$^2$, and width $\Gamma=172\pm 33$
MeV (statistical errors only). The $Y(4350)$ was confirmed by
BELLE\cite{belle_y4350}.

\section{The $X(3872)$}
\begin{figure}[htbp]
    \begin{center}
      {\includegraphics[scale=0.6]{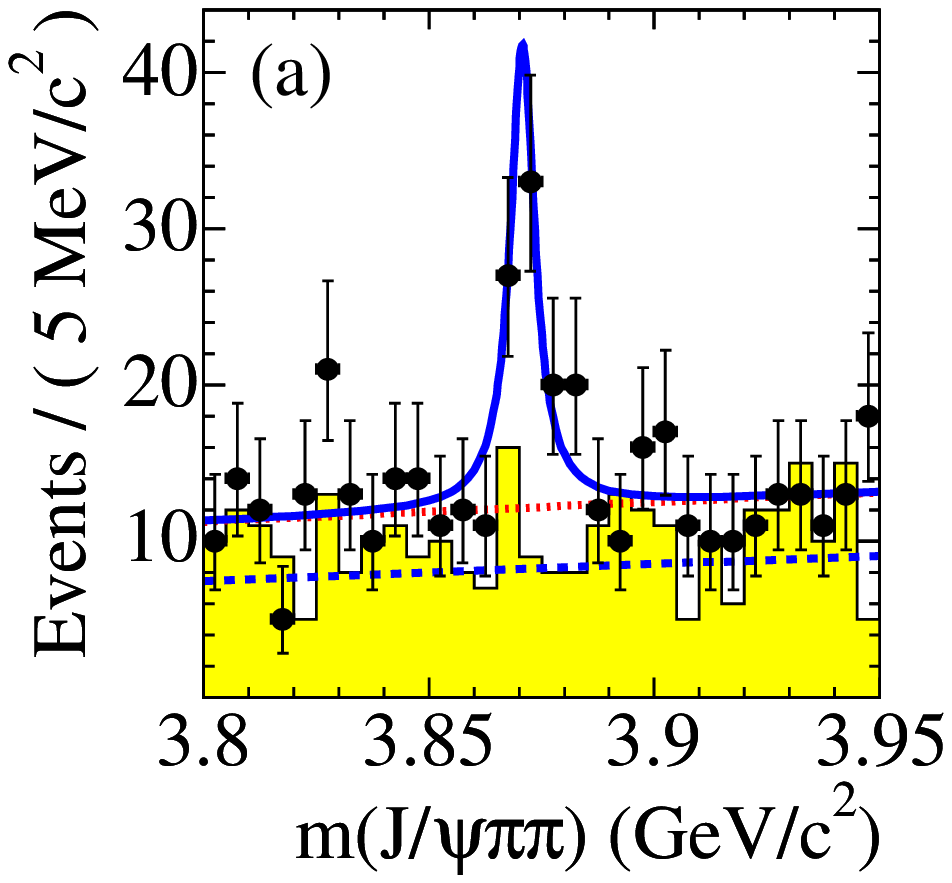}}
      {\includegraphics[scale=0.6]{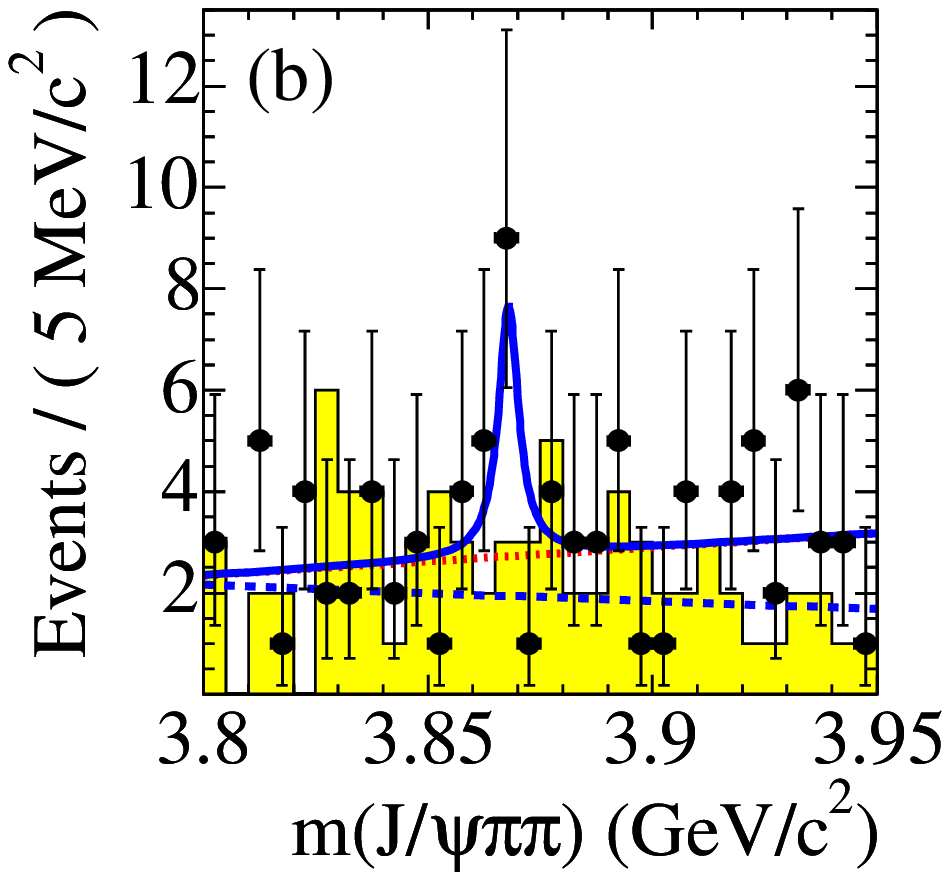}}
      \caption{\it The $J/\psi\pi^+\pi^-$ invariant mass distributions
	for charged (a) and neutral (b) $B$ decay from \babar\/. The
	dots represent the data, while the shaded histograms
	represent side-band background. The solid curves show the fit
	results, the dashed lines show combinatorial background, and
	the dotted lines represent the sum of combinatorial and
	peaking background contributions.}
      \label{x_3872}
    \end{center}
\end{figure}

The $X(3872)$ discovered by BELLE\cite{belle_x3872} was the first
of the new charmonium-like states. Later CDF\cite{cdf_x3872},
D0\cite{d0_x3872}, and \babar\/\cite{babar_x3872} confirmed the BELLE
observation. In \babar\/, a data sample of 211 fb$^{-1}$ was analyzed
to obtain the $J/\psi\pi^+\pi^-$ invariant mass in the region
$3.8-3.95$ GeV/c$^2$ separately for charged and neutral $B$-candidates
as shown in Fig.~\ref{x_3872}. The dots represent the data and
the shaded histograms represent background. For charged $B$ decay
(Fig.~\ref{x_3872}(a)), a clear enhancement is observed for
$m_{J/\psi\pi^+\pi^-}\sim 3870$ GeV/c$^2$. Statistically consistent
behavior is observed for the neutral mode (Fig.~\ref{x_3872}(b)).
The $X(3872)$ invariant mass obtained from the charged (neutral)
$B$-mode is $m=3871.3\pm0.6\pm0.1$ ($3868.6\pm1.2\pm0.2$) MeV/c$^2$,
and the corresponding branching fraction values are ${\cal BF}
(B^-\rightarrow X(J/\psi\pi^+\pi^-)K^-) = (10.1\pm2.5\pm1.0)\times
10^{-5}$ and ${\cal B} (B^0\rightarrow X(J/\psi\pi^+\pi^-)K^0) =
(5.1\pm2.8\pm0.7)\times 10^{-5} $ at $90\%$ C.L.

\begin{figure}[htbp]
    \begin{center}
      {\includegraphics[scale=0.5]{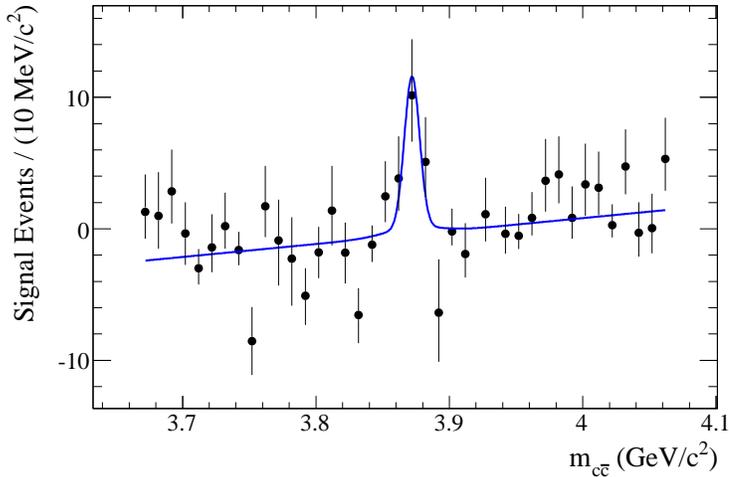}}
      \caption{\it The number of extracted signal events versus
      $m_{c\bar{c}}$ for the $X(3872)$ mass region. The solid curve
      represents the fit result.}
      \label{x_toJpsigamma}
    \end{center}
\end{figure}

In \babar\/, no evidence for a charged partner of the $X(3872)$ was
found\cite{babar_x0}, and so it is assumed that the $X(3872)$ has
$I=0$. Also \babar\/\cite{babar_xjpsig} has confirmed the BELLE
observation\cite{belle_xjpsig} of $X(3872)\rightarrow
J/\psi\gamma$. In Fig.~\ref{x_toJpsigamma} we show the $J/\psi\gamma$
mass distribution obtained from \babar\/, and a clear enhancement is
observed at the $X(3872)$ mass. It follows that the $X(3872)$ has
positive $C$-parity.

\begin{figure}[htbp]
    \begin{center}
      {\includegraphics[scale=0.45]{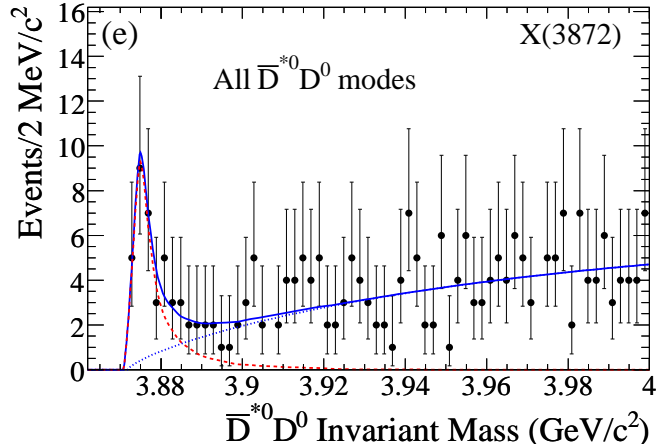}}
      \caption{\it The $\bar{D}^{\star0}D^0$ invariant mass, where the
	solid curve shows the fit result and the dashed curves
	represent the signal and background contributions.}
      \label{x_todd}
    \end{center}
\end{figure}

Later both BELLE\cite{belle_xdd} and \babar\/\cite{babar_xdd} have
found evidence for the decay mode $X(3872)\rightarrow
\bar{D}^{\star0}D^0$. In Fig.~\ref{x_todd}, we show the
$\bar{D}^{\star0}D^0$ invariant mass as reported by \babar\/. A clear
enhancement near threshold is observed. The measured mass values from
BELLE and \babar\ are $3875.2\pm0.7^{+0.9}_{-1.8}$ MeV/c$^2$ and
$3875.1^{+0.9}_{-0.7}\pm0.5$ MeV/c$^2$, respectively. The difference
between the mass value obtained in the $\bar{D}^{\star0}D^0$ decay
mode and that from PDG\cite{pdg} is then $3.8^{+1.2}_{-2.0}$ MeV/c$^2$
from BELLE and $3.7^{+1.1}_{-0.9}$ MeV/c$^2$ from \babar\/, indicating
that the effect is real. This mass difference has received a lot of
attention, although a simple explanation involving one unit of orbital
angular momentum (and hence $J^P=2^-$) has been proposed
recently\cite{dunwoodie}.

\section{The $Y(3940)$}
\begin{figure}[htbp]
    \begin{center}
      {\includegraphics[scale=0.5]{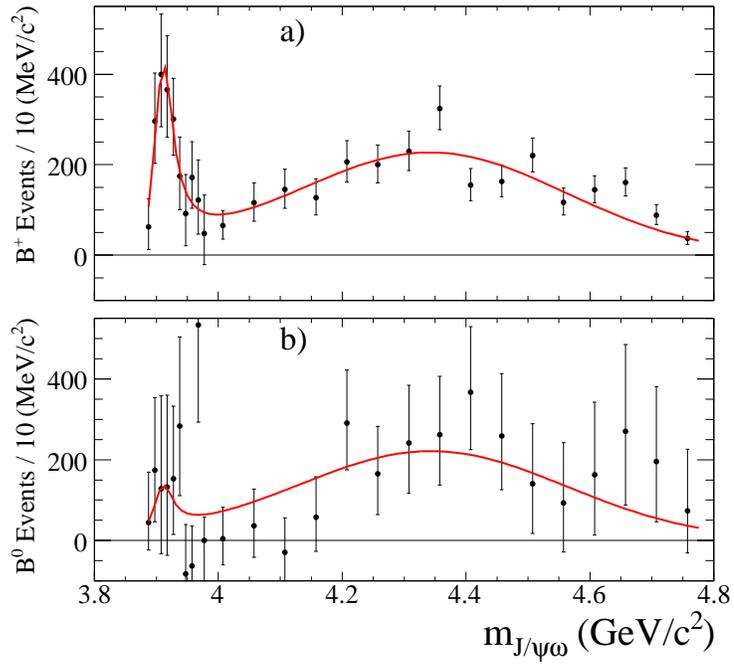}}
      \caption{\it The acceptance-corrected $J/\psi\omega$ mass
      distribution for (a) $B^+$ and (b) $B^0$ decay. The solid curves
      represent the fit results.}
      \label{y_3940}
    \end{center}
\end{figure}

The $Y(3940)$ was first observed by the BELLE
collaboration\cite{belle_y3940} in the decay process $B\rightarrow
J/\psi\omega K$ using 253 fb$^{-1}$ of data. The mass and width
obtained for this resonance were $m=3943\pm11\pm13$ MeV/c$^2$ and
$\Gamma=87\pm22\pm26$ MeV. In \babar\ a data sample of 348 fb$^{-1}$
is used to search for the $Y(3940)$\cite{babar_y3940}. Charged and the
neutral $B$-decays are analyzed separately, and in the \babar\
analysis finer mass binning was used on the basis of mass resolution
studies. Signal events were corrected for acceptance and mass
resolution effects. A significant enhancement is observed near
threshold in the charged mode, and a statistically-limited, but
consistent, signal is obtained in the neutral mode. In
Fig.~\ref{y_3940} we show the acceptance-corrected $J/\psi\omega$ mass
distributions for the charged ( Fig.~\ref{y_3940}(a)) and neutral (
Fig.~\ref{y_3940}(b)) $B$ decay modes, respectively. The data points are
fitted with a Breit-Wigner signal function and a single Gaussian
function for the non-resonant contribution. Good fits to the data are
obtained, as shown by the solid curves.
The mass and width of the $Y(3940)$ are found to be
$m=3914.6^{+3.8}_{-3.4}(stat) ^{+1.9}_{-1.9} (syst)$ MeV/c$^2$ and
$\Gamma = 33^{+12}_{-8}(stat)^{+5}_{-5}(syst)$ MeV, respectively, with
branching fractions for the charged and the neutral decay modes ${\cal
BF}(B^+\rightarrow YK^+)=(4.9^{+1.0}_{-1.0}(stat) ^{+0.5}_{-0.5}
(syst))\times 10^{-5}$, and ${\cal BF}(B^0\rightarrow YK^0) =
(1.5^{+1.4}_{-1.2}(stat)^{+0.2}_{-0.2}(syst))\times 10^{-5}$; the
latter has corresponding upper limit ($95\%$ C.L.) $3.9\times
10^{-5}$.

\section{Acknowledgments}
The \babar\ Collaboration acknowledges the extraordinary efforts of
the PEPII collider group for providing the excellent luminosity.  I
should like to thank the Hadron07 conference organizers for a nice
meeting at Frascati, and also Bill Dunwoodie and Walter Toki for
useful comments.


\begin{thebibliography}{99}
\bibitem{belle_x3872}S.-K.~Choi {\it et al.}, Phys. Rev. Lett. {\bf 91}, 262001 (2003)
\bibitem{cdf_x3872}D. Acosta {\it et al.}, Phys. Rev. Lett. {\bf 93}, 072001 (2004)
\bibitem{d0_x3872}V. M. Abazov {\it et al.}, Phys. Rev. Lett. {\bf 93}, 162002 (2004)
\bibitem{babar_x3872}B. Aubert {\it et al.}, Phys. Rev. {\bf D71}, 071103 (2005)
\bibitem{babar_y4260}B. Aubert {\it et al.}, Phys. Rev. Lett. {\bf 95}, 142001 (2005) 
\bibitem{babar_y4260phipipi}B. Aubert {\it et al.}, Phys. Rev.{\bf D 74}, 091103(R) (2006) 
\bibitem{babar_y4260dd}B. Aubert {\it et al.}, arXiv:0710.1371 [hep-ex], (2007) 
\bibitem{babar_y4260pp}B. Aubert {\it et al.}, Phys. Rev. {\bf D 73}, 012005 (2006)
\bibitem{babar_y4260B}B. Aubert {\it et al.}, Phys. Rev. {\bf D 73}, 011101 (2006)
\bibitem{babar_y4350}B. Aubert {\it et al.}, Phys. Rev. Lett. {\bf 98}, 212001 (2007)
\bibitem{belle_y4350}X. L. Wang {\it et al.}, Phys. Rev. Lett. {\bf 99}, 142002 (2007)
\bibitem{babar_x0}B. Aubert {\it et al.}, Phys. Rev. {\bf D 71}, 031501 (2005)
\bibitem{babar_xjpsig}B. Aubert {\it et al.}, Phys. Rev. {\bf D 74}, 071101 (2006)
\bibitem{belle_xjpsig}K. Abe {\it et al.}, arXiv:0505037 [hep-ex], (2005)
\bibitem{belle_xdd}G. Gokhroo {\it et al.}, Phys. Rev. Lett. {\bf 97}, 162002 (2006) 
\bibitem{babar_xdd}B. Aubert {\it et al.}, arXiv:0708.1565 [hep-ex], (2007)
\bibitem{pdg} W.-M. Yao {\it et al.}, J. Phys. {\bf G33}, 1 (2006)
\bibitem{dunwoodie}W. Dunwoodie and V. Ziegler, arXiv:0710.5191 [hep-ex], (2007)
\bibitem{belle_y3940}S.-K. Choi {\it et al.}, Phys. Rev. Lett. {\bf 94}, 182002 (2005) 
\bibitem{babar_y3940}B. Aubert {\it et al.}, arXiv:0711.2047 [hep-ex], (2007)

\end{thebibliography}
\end{document}